# Outflows from Evolved Stars: The Rapidly Changing Fingers of CRL618

Bruce Balick[1], Martín Huarte-Espinosa[2], Adam Frank[3], Thomas Gomez[3], Javier Alcolea[4], Romano L. M. Corradi[56], and Dejan Vinković[7]

ABSTRACT

Our ultimate goal is to probe the nature of the collimator of the outflows in the pre PN CRL618. CRL618 is uniquely suited for this purpose owing to its multiple, bright, and carefully studied finger-shaped outflows east and west of its nucleus. We compare new HST images to images in the same filters observed as much as 11 y previously to uncover large proper motions and surface brightness changes in its multiple finger-shaped outflows. The expansion age of the ensemble of fingers is close to 100y. We find strong brightness variations at the fingertips during the past decade. Deep IR images reveal a multiple ring-like structure of the surrounding medium into which the outflows propagate and interact. Tightly constrained three-dimensional ("3D") hydrodynamic models link the properties of the fingers to their possible formation histories. We incorporate previously published complementary information to discern whether each of the fingers of CRL618 are the results of steady, collimated outflows or a brief ejection event that launched a set of bullets about a century ago. Finally, we argue on various physical grounds that fingers of CRL618 are likely to be the result of a spray of clumps ejected at the nucleus of CRL618 since any mechanism that form a sustained set of unaligned jets is unprecedented.

## 1. INTRODUCTION

Pre-planetary nebulae ("pPNe") are formed from mass ejected in winds from the surface of an AGB star (Balick & Frank 2002). The traditional picture of AGB winds is that they are driven by isotropic radiation pressure absorbed on dust particles that form above their surfaces. However, Hubble images and molecular maps of pPNe show that the wind streamlines are far from isotropic. Indeed some AGB and many post-AGB outflows are highly organized and structured. This has led to a suggestion that the ejecta of an AGB star is somehow collimated by emerging magnetic fields at the stellar surface (Nordhaus et al. 2007) or by the influence of a binary companion star (De Marco 2009) that forms an excretion disk by mass overflow (Huarte-Espinosa et al. 2012a) and perhaps by an accretion disk on a companion (Soker & Kashi 2012). Mechanisms proposed to collimate the flows operate on such small size scales that direct observational tests of the nature of the engine are limited to the studies of the engine's large-scale exhaust plume. These tests have yet to be decisive. Thus even the general nature of the collimator remains a major challenge of stellar astrophysics after twenty years of active investigation.

[1] Department of Astronomy, University of Washington, Seattle, WA 98195-1580, USA; balick@uw.edu
[2] Department of Physics and Astronomy, University of Rochester, Rochester, NY 14627, USA; martinHE@pas.rochester.edu, afrank@pas.rochester.edu
[3] Astronomy Department, University of Texas, Austin, TX 78731-2330; gomezt@astro.as.utexas.edu
[4] Observatorio Astronòmico Nacional (IGN), E-28014 Madrid, Spain; j.alcolea@oan.es
[5] Instituto de Astrofísica de Canarias, E-38200 La Laguna, Tenerife, Spain; rcorradi@iac.es
[6] Departamento de Astrofísica, Universidad de La Laguna, E-38206 La Laguna, Tenerife, Spain
[7] Physics Department, University of Split, Teslina 12/III, HR-21000 Split, Croatia; vinkovic@pmfst.hr

This paper is an intensive case study of the structure and motions of the bright and well-studied outflows of the pPN CRL618. Its bright, shocked, ensembles of finger-shaped lobes provide an especially propitious opportunity to probe the mechanisms that may have formed them. At first glance it seems highly plausible that each ensemble is formed by some sort of "spray" of bullets ejected in a fairly recent event. The ensemble of (>20) hollow finger-like flows seen in $H_2$ and [FeII] found within the Orion Molecular Cloud, "OMC" (Allen & Burton, 1993) seems to be an interesting analogue. However, without constraining data other paradigms such as continuous jets have also been shown to be viable. Our high-level goal is to see if new multi-epoch images of CRL618 can be combined with other observations to determine whether one or more theoretical paradigms for the formation of fingers can be eliminated.

2-D hydrodynamical ("HD") simulations have been generally successful at matching the features of the fingers of CRL618, though not the ensemble of fingers as a whole. Lee & Sahai (2003, "LS03") and Lee, Hsu, & Sahai (2009, "LHS09") showed that the shapes and shocked optical emission of the fingers could be modeled by ongoing "diverging stellar winds" (that they denoted as "jets"). Then Dennis et al. (2008, "DC+08") showed that outflows consisting of bullets or their steady-state counterparts, "cylindrical jets", were also capable of explaining the observations then available. We shall present much more sophisticated but otherwise similar models later. NB: We shall use the term "jet" only in reference to cylindrical (non-diverging) flows.

The two classes of models differ in at least one important way. Diverging winds flow into and sweep the ambient gas inside of the fingers to their edges leaving a largely empty cavity filled only by the diverging winds. The majority of the outflowing mass lies in compressed wind and swept up AGB winds along the edges of the fingers which may converge to form bright fingertips. In contrast, cylindrical jets flow along the symmetry axis of the fingers and, like a thin piston, form a leading supersonic bow wave that, like a speed boat, pushes gas aside as the piston penetrates the ambient medium. The cavity contains the dense jet (or clump) and the low-density gas left from the backwash of the advancing bow wave. The total mass inside the cavity is essentially that of the displaced AGB wind plus the clump or thin jet. The densest material is in the clump or jet (though this may not be radiative); like the diverging winds the brightest emission comes from the shock-formed fingertips over a wide range of wind outflow speeds.

We return to the question of comparing model predictions with data in sections 4 and 5 after we present and discuss our new observational results (sections 2 and 3). First, however, we pause to review the current observational literature on this object.

*Summary description of CRL618.* CRL618 ($04^h42^m53^s.6 +36°06'53''$, PN G166.4–06.5) is one of the most intensively studied pPNe owing to its ensembles of multiple and bright high-speed fingers extending about 7″ east and west of its symmetry center. CRL618 was recognized as an unusual pPN shortly after its discovery by Westbrook et al. (1975). In 1976 Merrill and Stein (1976) suggested that its early-type central star is obscured within an inclined and dusty disk. CRL618 has turned out to be a member of a common subclass of pPNe that show bipolar structure along the symmetry axis of a central dust lane (Ueta et al. 2000; Sahai et al. 2011). Bright, multiple finger-like outflows are rare but not unique to

CRL618.  IRAS 19475+3119 (Hsu & Lee 2011), IRAS 19024+0044 (Sahai, Sánchez Contreras & Morris, 2005) and IRAS 16594-4656 (van de Steene et al. 2008) also have finger-like outflows.  Of these, CRL618 is unique for its array of bright optical emission lines and the diagnostic information that they provide.

Following the lead of Sánchez Contreras, Sahai, and Gil de Paz (2002,"SCSG02") and Sánchez Contreras & Sahai (2004, "SCS04"), we adopt a distance D = 0.9 kpc.  At this distance a proper motion of 0.″1 per decade corresponds to a transverse speed of 45 km s$^{-1}$. Alternate estimates of its distance D range from 0.9 to 2 kpc (SCS04, Goodrich,1991, Schmidt & Cohen 1981).

The central star is heavily reddened (Chiar et al. 1998) and almost invisible owing to dense local and foreground dust lying orthogonal to the symmetry axis of the cones of fingers. The innermost core is a compact HII region detected in Br γ (Latter et al. 1995) and the radio continuum (Martin-Pintado et al. 1988 and Kwok & Bignell 1984).  Therefore an early-type star—possibly a B0 or cool carbon-rich Wolf-Rayet star with circumnuclear Hα lines (SCSG02)—lies at the center.  Polarized dust-scattered nuclear light, both Balmer emission lines and continuum, can be traced throughout the inner 5″ of the nebula (Trammell, Dinerstein, & Goodrich 1993 "TDG93" and SCSG02).  The outflow speed of the scattered nuclear light is similar to that of the ionized gas in the fingers implying that the scattering dust and the gas are parts of the same dynamical system (SCSG02).

At least five distinct hollow fingers are very prominent, each with bright knotty fingertips in F631N ([OI]), F656N (Hα), F658N ([NII]), and F673N ([SII]) images (Ueta, Fong, and Meixner 2001; Trammell & Goodrich 2002, "TG02").  Another five or more similar knots are seen in projection at smaller radii.  The trailing edges of each of the eastern fingers show a ruffled appearance with a notably regular repetition period of ≈0″.5.  TG02 argue that each finger has a different inclination angle $i$ from its neighbors by up to about 25˚.  SCS04 estimated that the cones containing the ensembles of fingers are inclined by an angle $i$ = 32˚ from the distribution and kinematics of CO emission.  SCSG02 estimated 20˚ < $i$ < 39˚ from the dynamics of optical emission lines observed along the fingers.

The Doppler speeds of the fingertips are as high as ≈200 km s$^{-1}$ as measured from high-dispersion optical long-slit spectra (SCSG02).  TDG93 found that the receding west cone suffers considerably more foreground extinction (E(B-V)=2.1) than its eastern counterpart (E(B-V)=1.3).  They attribute this to a tilted equatorial disk that lies in front of the western fingers of CRL618 and behind the eastern ones. Extinction estimates by SCSG02 are larger.

Intrinsic emission from a variety of atomic and ionic lines arises in the tips and the nearby lobe edges (e.g., TDG93; TG02; SCSG02; Riera et al. 2011, "RR+11").  The line ratios exhibit the very clear signature of shock-excited gas.  RR+11 find good matches of the line ratios with predictions of planar shock models (Hartigan et al. 1987, 1994) characterized by shock speeds $v_s$ = 30 to 40 km s$^{-1}$.  In a few places where faint [OIII] is observed the best match to planar shock models is $v_s$ = 80 to 90 km s$^{-1}$.  SCSG02 find very similar results. High-dispersion spectroscopy (SCSG02) that shows that the Doppler widths in the bow shock reach 80 km s$^{-1}$ in the brightest of the fingers.

TG02 and RR+11 find that density in the fingertips is at least $10^{3.6}$ cm$^{-3}$ from the ratio of the [SII]6716, 6731Å lines. This implies that each of the [SII] lines can be quenched by collisional de-excitation. SCSG02 measure similar densities and show that the density along their slits is constant. RR+11 surmise that the electron density $n_e$ could be as high as $10^{5-6}$ cm$^{-3}$ from the ratio of these forbidden lines to Hα. TG02 and SCSG02 point out that the recombination time of the S$^+$ ion is of order $(10^4$ cm$^{-3})/n_e$ y. Thus the emission-line shocks in and near the fingertips must be continuously sustained by some sort of "piston" of high-speed gas originating from the nucleus.

The radio and infrared spectra of CRL618 show a large variety of bright carbon- and oxygen-rich molecules (Remijan et al. 2005). CO, HCO$^+$, and H$_2$ and their kinematics have been mapped (e.g., Kastner et al. 2001, "KW+01"; Cox et al. 2003, "CH+03"; SCS04; Nakashima et al. 2007, "NF+07"). SC+04 found that the brightest CO consists of several kinematic components, including three low-velocity components ("LVC"s; an equatorial torus, an extended polar feature adjacent to the cones containing the lobes, and an amorphous halo), and one high-velocity component ("HVC") extending ≈2″.5 along the base of the cones (also visible in other CO lines; e.g., NF+07). The HVC is characterized by a Doppler shift that rises from ≈20 to ≈170 km s$^{-1}$, a central density of $10^{6.7}$ cm$^{-3}$, a mass of 0.09 M$_\odot$, and an outflow age of 400 y. Similar molecular outflow speeds are reported by CH+03, SCSG02, and (Bujarrabal et al. 2010).

The dynamics of the molecular hydrogen were mapped along slits at PA = 90˚ (KW+01) and 94.5˚ (CH+03). The P-V diagram shows four components, all of them symmetric relative to the nucleus. These are: (1) either continuum or a broad nuclear H$_2$ line; (2) an LVC (±10 km s$^{-1}$) that can be traced to ≈ ±5″ E and W of the nucleus; (3) an HVC extending from the nucleus to ± 2″ at a maximum Doppler shift of 150 km s$^{-1}$; and (4) a second HVC extending along the fingers to ±5 or 6″. The Doppler shift of the second HVC H$_2$ rises monotonically from the nuclear zone to about ±150 km s$^{-1}$ near the fingertips, similar to the kinematics of stellar Hα reflected from dust in the fingers (SCSG02). These spatio-kinematic components can be identified respectively with (1) the circumstellar zone; (2) the slow AGB wind adjacent to the fingers; (3) the corresponding HVC seen in CO—possibly highly inclined fingers close to the nucleus; and (4) the obvious optical fingers extending ±7″ E and W of the nucleus.

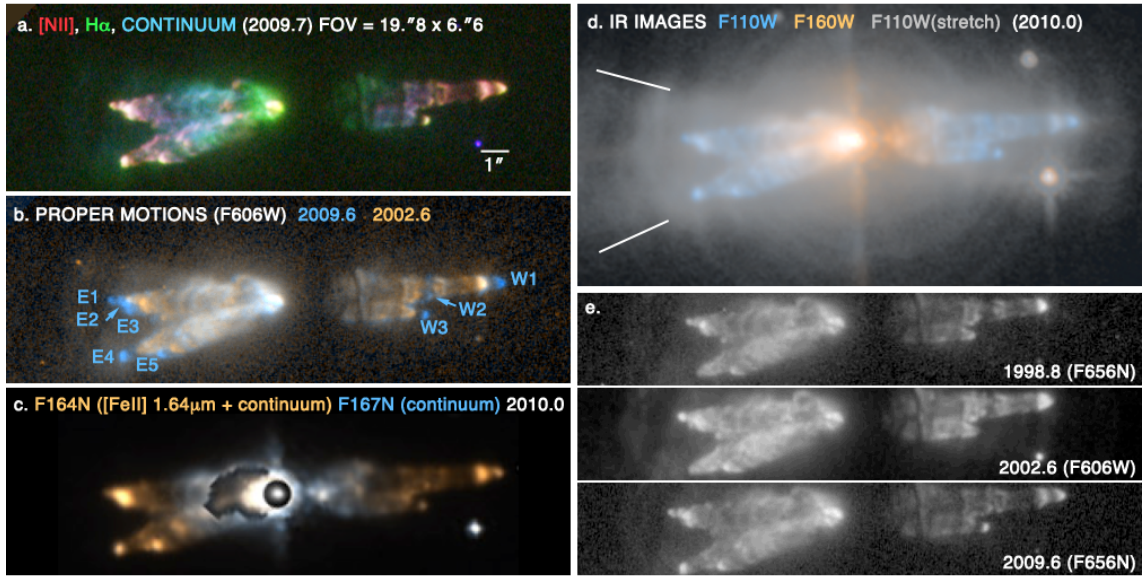

Fig 1. Images of CRL618  (a) Color overlay of F658N, F656N, and F547M images from 2009.7. Shock-excited lines originate in the pink regions near the fingertips.  Scattered light of the broad stellar Hα line dominates much of the green image layer.  Scattered continuum at 550 nm light appears blue.  (b) Overlay of a pair of F606W images from 2002.6 (orange) and 2009.6 (blue) showing the seven-year growth of the fingertips. (c) Overlay of F164N image (orange) and F167N (blue).  (d) Broadband IR images overlaid on a very-high-contrast rendition of the F110W image.  The edges of the eastern "searchlight" of scattered light is marked by a pair of white radial lines with opening angle 40°.  (e) Three images spanning 11 years that show the rapid changes in brightness of the fingertips, especially E1, E4, W1, and W3.

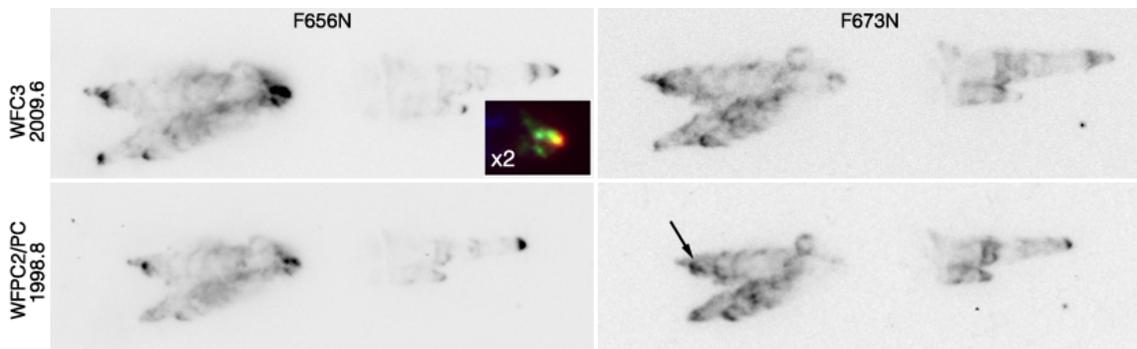

Fig. 2.  Linear inverse images of CRL618 from 2009.6 (top) and 1998.8 (bottom) that highlight changes of structure and brightness in the fingers and their tips in Hα (left) and [SII] (right).  All images are normalized so that fingertip E2 (arrow) is black. The field of view of each frame is 16″x 5″ (0.07 x 0.02 pc at D = 0.9 kpc).  The color inset is composed from F953N (red), F656N (green), and F673N (blue) images from 2009.7 and magnified by a factor of 2.  The peak of the central star appears as yellow.

| Camera | Pixel Size ″ | Filter | Exposure Time (s) | Observation Date | GO Program | Archive Data Set |
|--------|--------------|--------|-------------------|------------------|------------|------------------|
| WFPC2/PC | 0.046 | F656N | 1000 | 1998-10-23 | 6761 | U36A0105,6R |
| WFPC2/PC | 0.046 | F673N | 1200 | 1998-10-23 | 6761 | U36A0107,8R |
| ACS/HRC | 0.025 | F606W | 2240 | 2002-07-30 | 9430 | J6MS01010 |
| WFC3/UVIS1 | 0.0396 | F606W | 500 | 2009-08-07 | 11580 | IB1M02010 |
| WFC3/UVIS1 | 0.0396 | F547M | 520 | 2009-08-07 | 11580 | IB1M02020 |
| WFC3/UVIS1 | 0.0396 | F658N | 560 | 2009-08-07 | 11580 | IB1M02040 |
| WFC3/UVIS1 | 0.0396 | F656N | 560 | 2009-08-07 | 11580 | IB1M02050 |
| WFC3/UVIS1 | 0.0396 | F673N | 600 | 2009-08-07 | 11580 | IB1M02060 |
| WFC3/UVIS1 | 0.0396 | F953N | 135 | 2009-08-07 | 11580 | IB1M02030 |
| WFC3/IR | 0.128 | F110W | 1169 | 2010-01-12 | 11580 | IB1M06020 |
| WFC3/IR | 0.128 | F128N | 709 | 2010-01-12 | 11580 | IB1M06GJQ |
| WFC3/IR | 0.128 | F130N | 532 | 2010-01-12 | 11580 | IB1M06GKQ |
| WFC3/IR | 0.128 | F160W | 1169 | 2010-01-12 | 11580 | IB1M06010 |
| WFC3/IR | 0.128 | F164N | 790 | 2010-01-12 | 11580 | B1M06GLQ |
| WFC3/IR | 0.128 | F167N | 790 | 2010-01-12 | 11580 | B1M06GMQ |

Table 1.  Image Descriptions.

## 2. DATA AND ANALYSIS

We obtained images of CRL618 using WFC3 (Wide Field Camera 3 Instrument Handbook, version 2.1; Dressel et al. 2010) through visible-light filters F547M, F606W, F656N, F658N, F673N and near-IR filters F110W, F128N, F130N, F160W, F164N, and F167N (Table 1).  Detailed characteristics of the camera and its filters are described on the WFC3 Web sites.[8]  The visible-light exposures were limited to a single *HST* orbit in 2009 August.  These images, most of them dithered, have spatial FWHM of 0″.07 and were taken with a $512 \times 512$ subarray in order to expedite data transmission to the ground.  The IR exposures were obtained during another orbit in 2010 January using a 256x256 subarray.  Their PSFs are 0″.11 and 0″.16 at $\lambda = 1.1$ and 1.6 μm, respectively.  For reference, 0″.1 corresponds to 90 AU at the assumed distance of CRL 618.

All data were downloaded from the Hubble Legacy Archive.[9]  HLA images are bias subtracted, dark corrected, flat fielded corrected, converted to detected counts per second, and drizzled onto a uniform grid with north up.  Our subsequent data analysis methodology is described in an earlier paper (Balick et al. 2012, "Paper I").  Only the highlights are described below.  Please refer to Fig. 1.

*Color overlay of Fig. 1a.*  The color distribution of scattered light was measured from flux-calibrated WFC3 images in each filter after using bright field stars to align each image. In this figure (1a) the color saturation and contrast were altered in order to exaggerate the color variations. The variations of the ratios of our emission-line images are qualitatively consistent with existing optical spectroscopic observations.

[8] www.stsci.edu/hst/wfc3.

[9] Based on observations made with the NASA/ESA Hubble Space Telescope, and obtained from the Hubble Legacy Archive, which is a collaboration between the Space Telescope Science Institute (STScI/NASA), the Space Telescope European Coordinating Facility (ST-ECF/ESA) and the Canadian Astronomy Data Centre (CADC/NRC/CSA).  The relevant GO programs are listed in Table 1.  Support for program GO11580 was provided by NASA through a grant from the Space Telescope Science Institute, which is operated by the Association of Universities for Research in Astronomy, Inc., under NASA contract NAS 5-26555.

*Proper-motion studies*. One of the most prominent features of CRL 618 and a key finding of this paper is the prominent proper motions of the fingertips. These motions are seen immediately when the images are aligned using those few field stars in common to the images of each epoch. Figure 1b is an overlay of the F606W (broadband) images from 2002.7 (orange) and 2009.7 (cyan). The magnitude of their proper motions is easily discernable, as illustrated by the blue-orange displacements of E1 and W1. Figs. 1e and 2 show the highly conspicuous 11-year changes position and surface brightness of the fingers in the F606W, F656N, and F673 images.

Multi-epoch images were initially aligned using the two nearby field stars. Of course the field stars have individual proper motions. We found that features near the nucleus move when blinking multi-epoch images registered this way. Without other high-quality astrometric fiducials the alignment can only be improved by eye. Further north-south alignment is straightforward thanks to the sharp horizontal edges of the fingers. The final east-west alignment requires a very small translation until the perception of systematic motions of the knots near the center of CRL618 is nullified. These extra shifts have no substantial effect on the basic conclusions reached in this paper.

*Infrared (IR) images*. Fig. 1c and d shows color overlays of selected IR images. In the former case three linear brightness ranges are used to highlight the star, the nebulosity near it, and the fingers. The F164N filter transmits [FeII]$1.644\mu$m + Br (12-4)$1.641\mu$m. The F164N and F167N filters contain equal continuum contributions. Thus the orange regions show excess emission in the F164N filter. An IR spectrum of a point $2''.4$ E of the nucleus by HLD99 shows that Br (12-4) is not detectable. Therefore the fingers' tips and leading edges are outlined in [FeII] in Fig. 1c.

The F110W and F160W images are superimposed on the extended low-level background seen at $1.1\mu$m (shown in gray) in Fig 1d. The light in these images is largely scattered continuum that appears to peak at about $\lambda \approx 1\mu$m. An extensive system of rings surrounds the ensembles of fingers in the F110W image. The apparent size of the large-scale scattering system is limited both by its declining dust density and external illumination flux. The innermost ring-like features are verified in the broadband F160W image, albeit at slightly lower spatial resolution. Although CRL618 was previously observed at about 2 $\mu$m with NICMOS, the present images are the deepest and highest-resolution images of CRL618 in the near-IR spectrum.

*Fingertip brightness changes*. Figs. 1e and 2 highlight large changes in the optical brightnesses of the tips of the knots. The three frames are taken through two filters identified in the image and normalized so that the core is the same brightness in each frame. (The F606W frame appears generally brighter than F656N frames owing to its wide filter bandpass.) Note that a few fingertips temporarily disappear entirely in some of the images.

3. RESULTS

Unless noted otherwise, the results described below were extracted from a pair of F606W images obtained from the ACS in 2002.6 and WFC3 in 2009.6 since this image pair has the best spatial resolution, the cleanest point spread function ("PSF"), and highest S/N. The key findings derived directly from those images are these:

*Image colors.* Refer to Fig. 1a. The F547M image (shown in blue) is free of known bright emission lines. Thus the blue-colored "mid-finger" emission seen in this filter is dominated by scattered stellar continuum. As noted by TDG93 and SCSG02, the regions shown in green near the star are largely the result of scattered stellar H$\alpha$. (Oddly the scattered H$\alpha$ emission is relatively concentrated towards the nucleus.) The emission in the pink regions (a combination of [NII] and H$\alpha$ lines) is intrinic and arises from shocked gas. Knotty pink emission can be traced inside the fingers and along some of their edges. This is the signature of a bow shock. All of this has long been recognized. See RR+11 for details.

*Background features:* The extent and structure of the low-surface-brightness background is brightest and best defined in the F110W image (Fig. 1d). The background light arises in circular rings or ring fragments. The shapes and colors of the rings mimic those observed through the same filters in CRL2688 (Paper I) and resemble ensembles of rings found in many other pPNe and PNe. By analogy to CRL2688 (Paper I) and many other pPNe and PNe with similar sets of outer rings, the rings are most likely the result of mass-loss modulations in a geometrically diverging stellar wind originating long before the fingers were ejected. The gaps in the rings are likely to be regions of low dust density.

Two "searchlights" of scattered background light with an opening angle of about 40° from the nucleus are seen in the same general E-W direction as the fingers (Fig, 1d), also similar to CRL 2688. Clearly their geometry argue that the searchlights consist of nuclear light that escaped relatively unimpeded from the nuclear zone though regions that have been cleared of absorbing dust. It is difficult to measure the color of the illuminating light precisely since scattering changes its color; however, the searchlights are brightest in the F110W filter suggesting that their color temperature is about 3000K.

*Background expansion:* The proper motions of the rings are very small and difficult to measure owing to their faintness and diffuse structure. We find that magnifying the deep 2002.6 F606W image by a factor of 1.01 and blinking against its 2009.6 counterpart nulls any proper motion of the outer rings. The corresponding pattern expansion speed is about 20 km s$^{-1}$. However, magnification factors of 1.00 and 1.02 yield equally good results.

*Finger expansion ages and rates.* Using the pair of bright F606W images we find angular displacements 0″.46, 0″.45, and 0″.49 for fingertips E1, E4, and W1, respectively, relative to the position of the central star (Figs. 1b & 2). The corresponding kinematic ages are 128, 111, and 112y with 10% systematic uncertainties dominated by the definitions of the locations of the highly-variable fingertips. Similar results for the kinematic ages are found from F656N images spanning 10.8y.

*Overall expansion pattern.* The overall expansion pattern of the fingers was explored by blinking the F606W images after various magnification factors were applied to the earlier image. A magnification factor of $1.070 \pm 0.005$ nulls the motions of all of the bright outer fingertips when the two images are blinked. The pattern has an expansion age $t_{exp} = \theta \, \Delta t / \Delta \theta \sim 100 \pm 15$ y, independent of the distance to CRL618. Here $\theta$ is the radial displacement of the fingertip from the center of motion. The error of $t_{exp}$ accounts for other uncertainties. This agrees well with measurements of individual outer fingertips (above).

*Finger expansion speed*. At a presumed distance of 0.9 kpc the displacement of the longest fingertips W1 and E1 corresponds to 300 km s$^{-1}$ before any inclination correction. The proper motions of other fingertips are smaller in rough proportion to their nuclear separations. No proper motions are measureable for sharp features within 2″ of the nucleus in the F656W images within the uncertainties.

The measured proper motion compares to Doppler shifts of 80 km s$^{-1}$ found from long-slit Echelle spectra by SCSG02 and shock speeds $v_s$ of 40-80 km s$^{-1}$ reported by SCSG02 and RR+11 (see section 1). These fingertip speeds are considerably less than the speed of the proper motions. Thus the shocks that emit the observed optical lines are likely to arise in a reverse shock propagating into the dense gas that lags behind the true leading edge of the fingers. See section 4 for details.

Our [FeII] image (Fig. 1c) shows that the emission line arises at the tips and the leading edges of the fingers, just as it does in the fingers in the OMC cloud. In supernova remnants the line is created by shocks of several hundred km s$^{-1}$ (Graham. Wright, & Longmore 1989) in which iron-rich grains are sputtered. Thus the emission of [FeII] is consistent with the proper motions of the fingertips of CRL618 reported above.

*Fingertip brightness changes*. The knots at the tips of the fingers brighten and fade in optical images on time scales of a few years. Knots E4, W1, and W3 illustrate how radically the ends of the fingertips can brighten and fade on short time scales. In some cases the knot at the leading edge of the finger disappears entirely. These time scales are comparable to the cooling and recombination times of the shocks discussed in section 1. Although the projected locations of the rings are somewhat ambiguous, the fingertips W1 and E4 appear to brighten when the knots reach about the same radial offset as ring maxima fingertip (Fig. 1d). Therefore these fingers and the illuminated rings lie approximately in the same plane; that is, while the rings may be omnipresent, those that are seen in Fig. 1d lie in or near the searchlights that also contain the fingers.

*Fingertip structure*. A close inspection of the F547M, F606W, F656N, and F673 images from all epochs shows that the bright regions at the leading edges of the fingertips have a range of morphologies. The most common shape is a sharp-tipped "Vee" with the apex pointed away from the star (E3, E5, W1, W2, W3). In some cases the fingertip is a point or a small collection of point-like emitters (E1, E2). In a few other cases the fingertips are flatter – perhaps "Cee" shaped, especially in [SII] images shown in Fig. 2. The tips appear to retain their shapes as they propagate and vary in brightness. E4, W3 are good examples. Many fingertips show signs of substructure at the limit of the image resolution.

*Fingertip contrails*. Refer to Fig 1a and 1d. The ruffles seen in the "contrails" between the nucleus and the eastern fingertips have the same quasi periodicity as the adjacent reflection rings seen at 1.1μm (Fig. 1d). Indeed, they appear to align to the locations of the brighter rings next to them. The situation to the west (where the extinction is systematically higher) seems similar but is less clear. The growth rate of the ruffles, at most $0.1\%$ y$^{-1}$, is similar to the adjacent rings and much less than that of the fingertips, $1\%$ y$^{-1}$.

*Finger inclination angle*. The ratio of Doppler to proper velocities for the eastern fingertips is about 1/3 for a presumed distance of 0.9 kpc. The corresponding inclination angle $i$ is $20°$ from the plane of the sky. This estimate is very uncertain. For reasons discussed in section 1, we have adopted symmetry axis inclined at $i = 30°$ in our model computations in section 4.

*Possible circumnuclear disk*. A dark, lacy, largely vertical absorption feature is readily seen against the fingers extending out to $\approx 2''$ west of the nucleus of CRL618 (Figs. 1a & b) or possibly even further (SCSG02 Fig 10). Its extinction diminishes with wavelength and the emission behind it is largely visible in the F160W image (Fig. 1e). There is no counterpart to the east. This feature remains fixed when blinking the high-quality 2002.6 and 2009.6 images so its proper motion is indeterminate and its expansion age exceeds 500y. It seems likely that extinction arises in a tilted and nearly stationary nuclear disk in projection against the western fingers. We also note the presence of $H_2$ and CO features of the comparable size scale on both sides of the nucleus (KW+01, CH+03, SCS04). It seems probable that the molecular features and the foreground dust absorption feature are one and the same structure.

*Central unresolved nucleus*. The central "star" is visible in many of the optical and all of our IR images, though it is reddened to varying degrees by the dust lane to its west. Strong Balmer emission lines are found in its spectrum (TDG93). No stellar absorption features appear in its visible or IR spectra (TDG93, HLD99). The IR spectrum exhibits a strong red continuum and Paβ emission. The ratio of the stellar flux in F128N and F130N images is $\approx 5$ confirming that the nucleus emits bright Paβ.

Therefore the nuclear light probably arises in a dense, unresolved circumstellar shell of some sort, perhaps analogous to that of other PN and pPN nuclei with very bright nuclear Balmer lines such as M2-9, Hb12, M1-92, and many symbiotic stars.

*Nuclear brightening*. A comparison of two-epoch F656N, F606W, and F547M images in the archives shows that the nuclear source has brightened, at least relative to the two knots immediately to its east. The relative brightness change is about a factor of two in the F656N images. The F606W and F547M images show the same trend. Thus the intrinsic nuclear continuum is brightening, or the foreground extinction is decreasing, or both (see the two F673N images in Fig. 2). We add that the faintness of the nucleus and field stars and the proximity of the nucleus to much brighter knots render a precise quantitative measurement of the brightening impossible. Possibly related to this, SCS04 have noted that the radio thermal continuum flux at 2.3 GHz mm is variable.

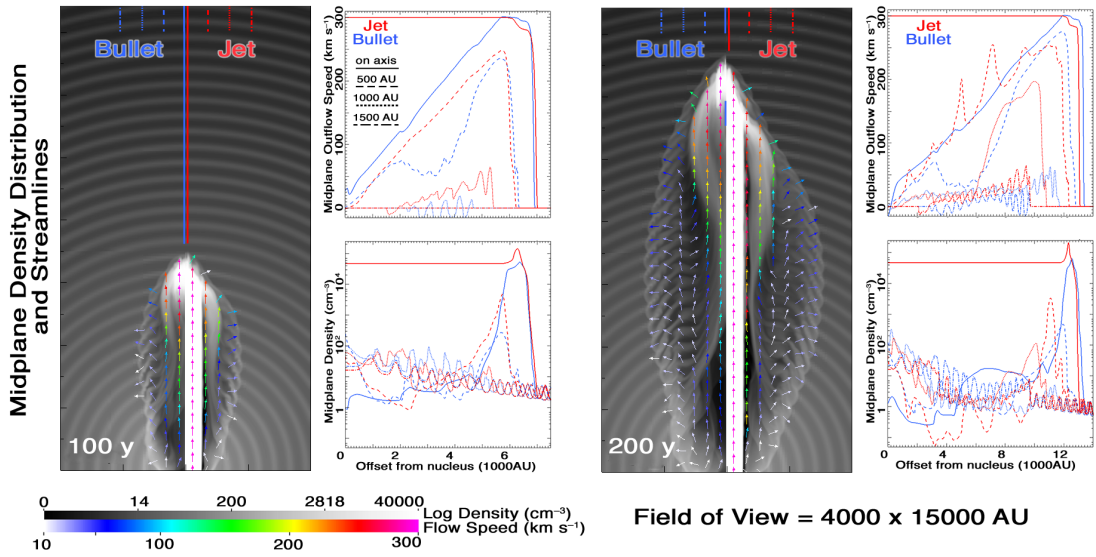

Fig. 3. Models of a bullet and jet with outflow speeds of 300 km s⁻¹ after 100 y (left half) and 200y (right half) propagating through an AGB wind with periodic density ridges spaced by 333 AU. The grey-scale panels show the density distribution and the streamlines in the midplane for bullets (left side) and jets (right side). The associated densities and radial velocities along lines in the midplane on or displaced from the symmetry axes are plotted to the right of each of the grey-scale images. The offsets are indicated by vertical solid, dashed, dotted, and dot-dashed tics along the top of each grey-scale image and the graphs to their right.

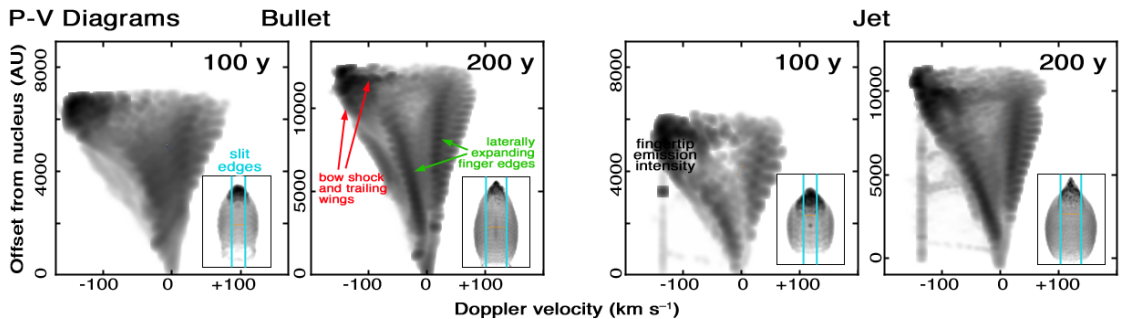

Fig 4. Predicted position-velocity diagrams of bullets (left pair of panels) and jets (right panels) at times t = 100y and 200y integrated over a slit. Slit widths are indicated on magnified 3-D model projections of the fingertips (insets). The symmetry axes of the P-V diagrams are inclined by 30° to the plane of the sky. The grey scales of the synthesized images (insets) are displayed logarithmically such that the peak brightness is black. Those of the P-V diagrams are greatly exaggerated to highlight the subtle features. In practice the light and neutral gray areas are not detectable. The images were generated using the program SHAPE (Steffen & Lopez 2006).

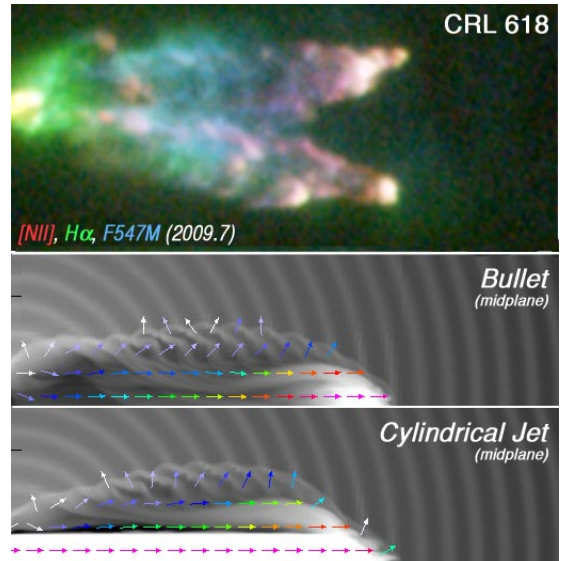

Fig 5. Bullet and jet model simulations of the midplane density and velocity vectors at 100y compared to the eastern ensemble of fingers of CRL618. In the top image the pink knots of Hα and [NII] emission are shock excited. Several other pink-colored knots are found throughout the ensemble.

## 4. COMPARISON TO MODEL SIMULATIONS

In this section we compare the results of our observations to HD models of pPNe including a set of new models described below. The models fall into two general classes: "ballistic" or "cylindrical" flows in which the out-streaming "fast" material from the star (sustained winds or bullets) all lies within a thin cylinder, and "diverging fast winds" where the fast stellar winds lie within a cone whose apex is an orifice at or near the wind source. In all cases the external environment is presumed to be a much slower, constant-speed wind from the AGB star whose large-scale density and ram pressure fall as $r^{-2}$.

### 4.1 Cylindrical Flow Models: Theory

Bullet and cylindrical jet models were first explored by Blondin, Fryxell, & Konigl (1990), Soker (2002), and DC+08), among many others. Cylindrical flows have several appealing properties. The mass and momentum of the fast winds create a piston that interacts in a narrow zone at the head of the flow and displaces the ambient AGB winds laterally in a bow wave. A fast shock forms at the forward edge, and a slower reverse shock propagates back into the piston. The bow wave at the leading edge and the outflow at the base of the jet generally have comparable speeds since the piston is too narrow to sweep up much of the gas upstream and decelerate. The length-to-width ratio of the lobes can be adjusted by specifying the forward speed of the piston and the sound speed of the AGB wind. The lobe edges are supersonic wakes that form as ambient gas is laterally displaced by the advancing piston.

Our models are Eulerian-grid numerical simulations that follow the formation of nebular lobes via the propagation of a bullet or a jet into an ambient AGB wind containing a set of mild bubble- or ridge-like density enhancements with spacings that correspond to the background ring separations seen in Fig. 1d. The equations of fluid dynamics are solved in 3 dimensions using the equations of hydrodynamics with radiative cooling. We use the methodology and the collision-dominated cooling curves described in previous papers, e.g., Huarte-Espinosa et al. (2012b). Radiative cooling rates are taken from Dalgarno & McCray (1972). The hydrodynamic equations were solved with the adaptive mesh refinement (AMR) numerical code AstroBEAR2.0[10]. In particular, the Euler equations with cooling source terms are solved using a second-order MUSCL Hancock shock capturing scheme and Marquina flux functions (Cunningham et al., 2009). We have ignored the effects of magnetic fields, gas self-gravity, and heat conduction.

The computational domain is a rectangle with dimensions $0 < x < 24000$AU, $-4000 < y,z < 4000$AU. We used a coarse grid with 300x100x100 cells along with two adaptive refinement levels to resolve the volumes in which the pressure gradients are large, increasing the grid resolution by a factor of 4. The maximum grid resolution is 20AU. Typical simulation flow times are of order 200y. We use Blue Hive[11] and Blue Gene/P[12] — IBM's parallel cluster and supercomputer, respectively—which are maintained by the Center for Integrated Research Computing of the University of Rochester. The present simulations ran for about 2 days using 256 processors.

Observational studies suggest that AGB winds expand isotropically with mass-loss rates and velocities $V$ of order $10^{-5}$ M$_\odot$ y$^{-1}$ and 20 km s$^{-1}$, respectively (see e.g. Hrivnak et al. 1989; Bujarrabal et al. 2001, "BC+01"). Accordingly the piston (bullet or jet) is launched at $V = 300$ km s$^{-1}$ corresponding to the proper motion of finger W1. The large-scale density profile, $n_{amb}(r)$, is a variation of Fig. 1 of NF+07:

$$n_{amb}(r) = (400\ cm^{-3})/(r/500AU)^{-2} \cdot 2(0.5 + \sum_i \exp(-[6\{r/500AU\} - 5/3 - i])) \qquad (1)$$

Here $i$ is an integer number that increments at each density ridge. The ridges are spaced every 500 AU. They were added to improve the fit of the model to HST images. The AGB wind speed is negligible compared to the fast wind, so for computational simplicity we allow it to be static. The ejected material is assumed to be too cold to increase its cross section owing to its thermal pressure or to cool radiatively.

We base our bullet model on the parameters adopted by DC+08. A spherical bullet moving at 300 km s$^{-1}$ was placed in the grid at coordinates $(2r_b, 0, 0)$, where the piston radius $r_b = 500$AU. The density profile of the bullet is given by

$$n_b(r) = (40000\ cm^{-3})\ (1 - [(x - 2r_b)^2 + y^2 + z^2]/r_b^2), \qquad (2)$$

which yields an initial bullet mass of $2.4 \times 10^{-5}$ M$_\odot$. See Yirak, Frank, and Cunningham (2010, "YFC10") for a detailed description of hydrodynamic simulations of bullets similar to those used here.

---

[10] https://clover.pas.rochester.edu/trac/astrobear/wiki

[11] http://www.circ.rochester.edu/wiki/index.php/BlueHive_Cluster

[12] http://www.circ.rochester.edu/wiki/index.php/Blue_Gene/P

For the jet model, we continually inject gas at a density of 40000 cm$^{-3}$ along parallel streamlines through a circular grid of cells located at the bottom face of our computational domain. The radius of the jet orifice is $r_j$ = 500A.U. and 2.2x10$^{-5}$ M$_\odot$ y$^{-1}$ flows through it at 300 km s$^{-1}$.

We remind the reader that it is expected that proper motions speeds, 300 km s$^{-1}$, will be far larger than the shock speeds inferred from emission lines. All models such as ours show that there are two shocks, one at the forward edge of the interaction zone, with speed $v_{fs}$, where gas compressed by the piston is thrust aside, and a reverse shock that moves upstream, with a speed $v_{rs}$, into the piston. In our case the speed of the reverse shock, $v_{rs}$ is given by

$$v_{fs} = v_{rs} \; \varLambda^{-1/2}, \qquad (3)$$

where $v_{fs}$ is the speed of the forward shock and $\varLambda$ is the bullet-to-ambient density contrast (Klein, McKee, & Colella 1994; YFC10). The actual density contrast of the bullets of CRL618 is not measureable directly. The measured shock speeds range from 40-80 km s$^{-1}$, a result that is consistent with $\varLambda \approx 30$.

## 4.2 Cylindrical Flow Models: Results

We present the results of our model simulations in the computational midplane in Fig 3. Columns 1 (3), show the 2D density and velocity vectors at an elapsed time $t$ = 100 y (200 y). In these maps the bullet and the jet models are displayed side by side. The bullets and jets lie within the white regions (density = 40000 cm$^{-3}$) on the symmetry axis. Columns 2 (4) show 1D profiles of velocity (top) and density (bottom) at these times. The small oscillations in density in many of the profiles correspond to the ridges in the AGB wind profile.

In Fig. 4 we present predicted slit-summed position-velocity ("P–V") diagrams and projected optical emission-line images of the bullet and the jet models at $t$ = 100 y and 200 y. The contrast of the diagrams has been greatly exaggerated to highlight any differences among them.

*Model Similarities*. Bullet and jet models share many similarities and, for the most part, these results are all nicely consistent with the observations. The bow shocks of both propagate at the ejection speed, 300 km s$^{-1}$, away from the densest region (upwards) of the ambient media, forming an elongated trailing lobe, or finger. The piston deflects ambient gas laterally forming a wake along the edge of the lobe. The morphologies and speeds of the bow shocks and the lobes are nearly the same.

The trailing finger edge (i.e., wake) expands laterally as the finger grows in length. The finger is nearly isomorphic no matter whether the piston is a bullet or cylindrical jet. Because of the contact discontinuity ("CD") along the finger perimeter the external AGB wind does not mix with gas inside the lobe (comprised of material that was shed from the piston). Thus AGB wind cannot penetrate the walls of the fingers. Lines of H$_2$ can be excited by the relatively low-speed shocks along the edge. The speeds of the edge shocks increase with distance along the outflow axis. The speeds are sufficient to excite bright (weak) lines of [NII] (Fig. 1a), [SII] (Fig. 2), and [FeII] (Fig. 1c) at the finger tip (lateral edge).

Regularly separated vertebrae-looking features are shed along the lobe walls as the bullet and the jet encounter the ambient ridges in the AGB density. The axial velocity of these features decreases with radial distance from the axis, as do their density (below). The synthetic emission (insets in Fig. 4) associated with theses features resembles the observed "ruffles" along the fingers' edges (see Figs. 1 and 2). This is the first time that such features have been simulated in numerical models of situations of this type.

Bright fingertips of both bullets and jets are highly ephemeral in our models. The emission flux from the shocks lying in fingertips decreases secularly as the density of the downstream AGB wind declines (affecting the compression and emissivity of the shocks) and as the detailed geometry of the head of the piston evolves. We find that the emission from jets drops in brightness by roughly a factor of 100 every 100 years for the first 200 y. This is largely caused by the lower inertial pressure of the AGB winds at large radii. The brightness of bullets drops even faster as ablation decreases their cross sections.

On much shorter times scales emission from the fingertips temporarily intensifies by an order of magnitude as the bullets or jets encounter the crests of the ridges in the AGB winds. The amplitude of the brightness variations scales with the square of the varying density contrast across the shock as the piston travels from peak to trough.

*Model Differences.* Next we look for differences in bullet and jet models that might help to distinguish one from the other observationally. The only major differences arise along the symmetry axis, or spine, of the lobes. By assumption the density and the outflow speed of jets is large and uniform on the axis. In contrast the corresponding volume behind bullets is much lower in density and the gas velocity drops linearly with distance behind the bow shock. Therefore it is easy to distinguish between bullets and jets—provided, of course, that the gas in this volume radiates. In our images (Fig. 5) the fingers appear largely hollow.

There are some other smaller differences in the predictions of bullet and jet models. The cross section of the jet remains fixed as the width of the lobe behind it increases. The bullets shed their mass largely from their sides. So their cross section and total mass decrease with time (see also YFK10 Figs. 3 & 4 in which the clumps are assumed to have sharp edges). Monitoring the ratio of jet (or bullet) and finger width is beyond the capability of HST images. On the other hand we find that the bullet breaks apart shortly after 200 y and the fingertips rapidly fade. The breakup will be observable when it occurs.

In summary, the hydro models show that the shapes and kinematics of fingers produced by bullets and cylindrical jets are all but observationally indistinguishable, at least when observed in shock-excited lines that arise at their tips and along their edges. (Here we assume that the cylindrical jets or bullets are not directly observable.) In this case a lobe formed by a bullet will show a nearly linear correlation of speed and distance along the symmetry axis whereas a jet will show a constant outflow speed (compare the solid red and blue lines in the upper graphs of Fig. 4). The kinematic patterns also differ somewhat along a thin sheath surrounding the flow column. In the case of CRL618 the spatial resolution of the kinematic observations must be $\leq 0''.1$ for the kinematic differences of the jet and bullet models to become apparent.

### 4.3. Diverging-Wind Models

This brings us to the next finger-forming paradigm: diverging (or conical) fast winds. Cantó, Tenorio-Tagle & Rozyczka (1988) first used diverging-wind models to explore H-H outflows. The paradigm was applied to PNe by Frank et al. 1996). As noted in section 1, diverging flows interact with AGB winds qualitatively differently than do cylindrical flows. The diverging winds sweep out hollow lobes. Eventually the winds obliquely ram into and shock the inner edges of the thin rim (or shell) of compressed gas. The radial momentum transferred to the rim causes it to expand supersonically in the ambient AGB material. A second (leading) shock forms on the outside edge of this rim as the rim is pushed into the AGB winds. A CD isolates the fast wind from the AGB material along the finger edges. The fingertips advance into the AGB winds at speeds at about half of the initial speed of the fast wind as it first enters the lobe.

LS03 and LHS09 constructed models of the interactions of fast winds that diverge into a narrow cone and flow into the AGB winds. Their model computations are based on ZEUS 2D (no adaptive grid), but are otherwise similar to ours. LS03 ran a series of models in which parameters were adjusted to account for the finger shapes, expansion speeds, and observed morphology and excitation of the shock emission seen optically. At the time when they made their models (before 2003) the fingertip speeds were thought to be 150 km s$^{-1}$. Accordingly their models include combinations of winds speeds of 300 and 1000 km s$^{-1}$, a cone opening angle of 10 and 20°, and steady or pulsed winds. Very similar model parameters were used by LSH09 to explore radio and IR molecular line observations that had appeared by 2007.

The effort of LS03 and LHS09 to match this wide range of observations to models was a huge and largely successful undertaking that we cannot review in detail here. The model with steady winds of 300 km s$^{-1}$ and a 10° opening angle gave the best overall results to the nebular morphology and observed shock-excited spectrum. The velocity pattern of the streamlines flowing within the walls of the cavity increases almost linearly with distance along the symmetry axis. LS03 generated synthetic P-V diagrams that are observationally indistinguishable from those of cylindrical jets and bullets in Fig. 4. In summary, and with very few exceptions, the observables predicted by cylindrical and diverging wind models of CRL618 discussed above are extremely similar.

### 4.4. Applicability of the Models

Rather than dwell on their successes, which are many, we highlight the few issues that were not resolved by the diverging-wind models of LS03 and LHS09. Firstly, the predicted speed of the fingertips of CRL618, 150 km s$^{-1}$, in diverging-wind models is only about half of the proper motion speed observed in repeated HST images (section 3). This is probably of no consequence since this was the outflow speed known at the time of they developed their models. (A faster wind speed could probably be found to match the observations of CRL618 unless cooling behind the shocks becomes ineffective.) Secondly, the shock speeds at the fingertip derived from the diverging-wind model are too slow to explain the optical line ratios observed there. Again, a faster initial wind speed will help.

A third and potentially more serious problem for the LS03 models is that they assumed that the structure of the AGB wind is smooth. However, the ridges that are now seen in the AGB wind (Fig. 1d) are likely to deflect and defocus the flow within the thin walls of the fingers that would otherwise converge at the fingertips. This is simply conjecture on our part. Realistic and fully 3D hydro models with an adaptive grid meshes are needed to check the stability of flows in the walls of an AGB wind with structural inhomogeneities.

Recall that the optical emission lines are brightest in fast shocks ($v_s > 50$ km s$^{-1}$) and that all types of models can readily explain their properties. Therefore it is instructive to compare the models to the nebular structure revealed by $H_2$ observations (section 1) since $H_2$ is a tracer of relatively slow-speed shocks and extinction is less of a problem. The excitation of $H_2$ takes place in shocks of speed greater than 5 km s$^{-1}$ but less than 25-45 km s$^{-1}$ that dissociate the $H_2$ (Kwan et al. 1977, Draine, Roberge, & Dalgarno 1983).

Therefore the $H_2$ LVC ($H_2$ component 2) almost certainly originates in the AGB wind that has been plowed and displaced by the growing lobes. The Doppler shifts of the $H_2$ lines are easily explained if the lines arise in the displaced, laterally flowing AGB material along the outer edges of the fingers. This is also the pattern observed along the edges of the fingers behind the OMC bullets (see the subarcsecond image of $H_2$ and [FeII] at http://www.gemini.edu/index.php?q=node/226).

As noted in section 1, the Doppler shifts of $H_2$ in the HVCs can be traced to ±150 km s$^{-1}$ (KW+01 Fig. 6, CH+03 Figs. 4, 5). Because the $H_2$ would be quickly dissociated in shocks of speeds $v_s > 50$ km s$^{-1}$, the $H_2$ lines in the two HVCs presumably arise in volumes that are distinct from those seen in optical lines, *to wit*. high-speed gas interior to the fingers or along their optically bright lateral edges. This is a key point. In diverging-wind models all of the gas at high speeds lies within the dense walls of the lobes and have speeds that increase with distance from the nucleus. However, the gas streamlines first cross through fast shocks on the inner edges of these walls where shock speeds dissociate the $H_2$. (The LS09 models also show that gas within the walls is also subject to strong shears.)

The cylindrical flow models do slightly better at explaining the presence of the $H_2$ HVCs. The flow pattern of the gas in the interiors of the lobes (but outside of any cylindrical jet) rises secularly with offset from the base of the lobes (cf. the solid blue and dashed red lines in the upper graphs of Fig. 3) as the $H_2$ data require. However, it remains to be seen whether the small column density of molecular hydrogen inside the lobes is compatible with the observed brightness of the $H_2$ line emission.

To summarize, all of the HD models that were optimized to explain individual fingers of CRL618 lead to sets of outcomes that agree well with one another and with most of the data available. On the whole, the cylindrical-flow models encounter fewer problems in explaining the Doppler patterns revealed by $H_2$ observations than do diverging-flow models. Higher-resolution (≤0″.5) kinematic mapping of molecular lines in the interiors of each of the fingers would be extremely helpful[13].

---

[13]   Adaptive-optics (AO) imaging of $H_2$ in CRL618 requires the availability of a close and bright star for phase reference. Unfortunately the nearest such star is almost 2′ distant.

Finally, it should be emphasized that the models are strictly phenomenological, not comprehensive.  None of them account for the formation of fast winds at the base of the fingers, the collimation process, or the conspicuous ensembles of multiple fingers in CRL618, all with similar ages.

## 5. SUMMARY AND CONCLUSIONS

*Summary of the observations and the models.* In this paper we have used new optical and IR images of CRL618 obtained from HST to explore the nature of its fingers of highly collimated outflows. Additionally we have used these and other observations to constrain and evaluate two types of credible models of their origins. Our major observational findings are:

The most extended fingers grew by $\approx 0''.5$ in seven years. At a presumed distance of 0.9 kpc this proper angular motion corresponds to 300 km s$^{-1}$. The proper motions of shorter fingers are proportionally smaller. This compares to measured Doppler shifts of $\approx 80$ km s$^{-1}$ for optical lines and $\approx 150$ km s$^{-1}$ for CO(2–1) and $H_2$ found in the literature.

The nebular growth is largely homologous. All fingers have the same expansion ages, $\sim 100$ y, independent of presumed distance to CRL618.

There is no direct evidence of high-speed jets, diverging winds, or bullets within each finger, though we can confidently surmise from the presence of bright bow shocks that one of these types of outflows applies.

The ruffled "contrails" in the wake of the fingertips have the same quasi periodicity as the surrounding rings seen in dust-scattered starlight at 1.1μm (Fig. 1a and d). The ruffles lag substantially behind the overall expansion pattern of the fingertips. They seem cospatial to and affected by the surrounding ridges of dusty, cold AGB winds.

The leading edges of the most rapidly-moving fingertips show a Vee geometry that point to an underlying piston of unresolved cross section. Other fingers have more complex shapes.

The leading edges of each of the fingers brighten and fade on time scales of a few years. The brightening appears to occur when the changing locations of the knots coincide with peaks of scattered starlight in the ridges of ambient AGB-wind (Fig. 1a and e) and the fading occurs where the scattered light is faintest.

The leading edges of the fingers are easily visible in [FeII] lines at 1.64μm. This line arises in fast shocks where iron grains are sputtered. The distribution of [FeII] emission complements that of $H_2$. That is, [FeII] is strongest near the fingertips and $H_2$ is brightest closer to the nucleus. (Much the same pattern is seen in the OMC fingers.)

To better understand these essential behaviors we computed high-resolution 3-D hydro-dynamic simulations of dense bullets and thin cylindrical jets as they encounter and displace ambient AGB winds containing circular ridges of high density. We showed that these models fit the observations quite nicely and explain the variability in the brightness of the fingertips. The fits to the lobe morphologies, growth patterns, and kinematics are very comparable to previous models of LS03 and LHS09 in which fast, diverging winds sweep out lobe interiors, deflect along the fingers' walls, and then converge to form the bright, shocked fingertips. The principal results of the data-model comparisons are these:

Like the diverging wind models of LS03 and LHS09, both jet and bullet models produce finger-like lobes whose morphologies are in generally good agreement with extant imaging observations. All pistons at 300 km s$^{-1}$, whether jets or clumps, create slower reverse shocks that account for the properties of optical emission lines. The bullet and cylindrical jet models can explain the increasing Doppler shift of the $H_2$ with nuclear offset if the emitting $H_2$ fills their interiors. As LHS09 noted, the kinematic images of $H_2$ by CH+03 are not consistent with predictions of the diverging wind models.

The ruffled edges of the fingers seen in HST images are intimately connected to the relatively high-density ridges in the ambient AGB material. Both cylindrical-flow models, bullets and jets, explain how the ruffled edges arise as the fingers evolve. Numerical models of diverging-winds are needed to determine whether dense ridges of AGB wind will scatter the wind streamlines before they converge at the fingertips.

The preponderance of evidence of extant observations favors the thin-jets or bullets. Bullets are also favored by the pattern of steadily increasing $H_2$ kinematics and by analogy to other outflows such as the ensembles of finger-shaped outflows in the Orion Molecular Cloud.

*The distinction between jets and bullets.* As noted in section 1, the nature of the outflows that have formed the ensembles of multiple fingers in CRL618 has broad implications for stellar evolution. If we can distinguish bullet and jet models then we can infer whether the process that form them act briefly or systemically. Until we make this distinction even the broad outlines of the formation processes of fingers and similar lobes are unclear. The observations are not decisive. We end the paper with a brief discussion of which class of outflows seems most plausible on other grounds.

There are two strong reasons to prefer the bullet hypothesis. The first is that any mechanism that can form simultaneous but unaligned jets is difficult to imagine. The most probable mechanism for jet formation in other types of objects is some form of magneto-rotational launching (Blandford & Payne 1982; Ouyed & Pudritz 1997; Blackman, Frank, & Welch, 2001; Mohamed & Podsiadlowski 2007). This class of model, and one that applies very generally to jets from YSOs and AGNs, requires a rotating, strongly magnetized central source such as a star surrounded by an accretion disk. The magnetic fields serve as a "drive belt" for converting rotational energy of the disk into kinetic energy of the outflowing material (though in some cases the "piston" driving material is actually Poynting Flux"). Driving several steady (or semi-steady) simultaneous jets in different directions requires an entirely new form of jet launching that isn't the result of disk rotation or typical magnetic fields.

On the other hand, the ejection of a series of roughly bipolar, coeval bullets or clumps is much easier to understand. There are numerous ways to launch a spray of bullets from rapid or asymmetric ejections of dense matter in which instabilities and accelerating flows play an early role. For example, Blackman et al. (2001) and Matt et al. (2006) proposed the concept of a "magnetic bomb" to explain the collimation of stellar outflows in AGB stars. The mechanism supposes that the build-up of subsurface toroidal field overwhelms the gravitational weight of overlaying layers leading to a rapid and very turbulent bipolar ejection of those layers. Recently Aksahi & Soker (2013) proposed that a jet launched from a nearby compact companion can intercept and accelerate segments of dense ejected mass shells from an AGB star (Aksahi & Soker 2013, preprint). Under the right conditions this can lead to the ejection of multiple high-speed clumps formed from R-T instabilities within the shell.

In addition, laboratory studies have shown that episodic Poynting Flux Dominated outflows naturally fragment into a chain of collimated clumps via kink mode MHD instabilities (Lebedev et al. 2005). Astrophysical simulations of this process confirm that the brief ejection process becomes unstable at the height of its activity and produces a spray of clumps with similar speeds, directions, and kinematic ages (e.g., Huarte-Espinosa et al. 2012a).

A second but much less compelling reason to prefer bullets is that the old and new images of CRL618 are scaled versions of one another. The fingers of CRL618 are straight and show no sign of a precessing or varying jet launcher. They could easily be the ballistic shrapnel of ejections with a common age.

In a future paper we will reconsider the applicability bullet/jet paradigms when we analyze images of other pPNe with highly symmetric biconical outflows.

*Acknowledgements:* A special thanks to the referee for careful, thoughtful, and insightful suggestions on the initial version of this paper. Garrelt Mellema provided very thoughtful and insightful comments abut the workings and predictions of hydro models. We also thank Max Mutchler of STScI for help with the image calibrations from different HST cameras. The hospitality of Stockholm Observatory (where most of this paper was written) is greatly appreciated by B.B.

This project was supported by HST GO grant 11580. Support for GO11580 was provided by NASA through a grant from the Space Telescope Science Institute, which is operated by the Association of Universities for Research in Astronomy, Incorporated, under NASA contract NAS5-26555. T.G. is grateful for support from a Boeing Scholarship and a McNair Fellowship awarded through the office of Undergraduate Academic Affairs at the University of Washington. J.A. is partially supported by the Spanish MICINN, program CONSOLIDER INGENIO 2010, grant "ASTROMOL" (CSD2009-00038).

Some of the data presented in this paper were obtained from the Multimission Archive (MAST) at the Space Telescope Science Institute (STScI). STScI is operated by the Association of Universities for Research in Astronomy, Inc., under NASA contract NAS5-26555. Support for MAST for non-HST data is provided by the NASA Office of Space Science via grant NAG5-7584 and by other grants and contracts.

Facility: HST (WFC3)